\documentclass[10pt,conference]{IEEEtran}

\usepackage{cite}
\usepackage{amsmath,amssymb,amsfonts}
\usepackage{graphicx}
\usepackage{textcomp}
\usepackage{xcolor}
\usepackage[hyphens]{url}
\usepackage{enumitem}
\usepackage{xspace}
\usepackage{booktabs}
\usepackage[listings]{tcolorbox}
\usepackage{array}
\usepackage{tabularx}
\usepackage{macros}
\usepackage[hyperfootnotes=false]{hyperref}

\renewcommand{\footnoterule}{%
  \kern -3pt
  \hrule width 1.0\columnwidth
  \kern 2.6pt
}

\title{\sysname{}: An Agent-Driven Simulation Framework for LLM Serving Systems}

\author{
\begin{tabular}{
c@{\hspace{2.5em}}
c@{\hspace{2.5em}}
c
}
\begin{minipage}{0.27\textwidth}
\centering
Wonung Kim\textsuperscript{\textdagger}\\
\textit{KAIST}\\
Daejeon, Republic of Korea\\
\href{mailto:wukim@casys.kaist.ac.kr}{\textcolor{blue}{wukim@casys.kaist.ac.kr}}
\end{minipage}
&
\begin{minipage}{0.27\textwidth}
\centering
Hyunmin Choi\textsuperscript{\textdagger}\\
\textit{KAIST}\\
Daejeon, Republic of Korea\\
\href{mailto:hmchoi@casys.kaist.ac.kr}{\textcolor{blue}{hmchoi@casys.kaist.ac.kr}}
\end{minipage}
&
\begin{minipage}{0.27\textwidth}
\centering
Minsu Kim\\
\textit{KAIST}\\
Daejeon, Republic of Korea\\
\href{mailto:mskim@casys.kaist.ac.kr}{\textcolor{blue}{mskim@casys.kaist.ac.kr}}
\end{minipage}
\\[3em]
\begin{minipage}{0.27\textwidth}
\centering
Jaehong Cho\\
\textit{KAIST}\\
Daejeon, Republic of Korea\\
\href{mailto:jhcho@casys.kaist.ac.kr}{\textcolor{blue}{jhcho@casys.kaist.ac.kr}}
\end{minipage}
&
\begin{minipage}{0.27\textwidth}
\centering
Yeongwook Kim\\
\textit{KAIST}\\
Daejeon, Republic of Korea\\
\href{mailto:ywkim@casys.kaist.ac.kr}{\textcolor{blue}{ywkim@casys.kaist.ac.kr}}
\end{minipage}
&
\begin{minipage}{0.27\textwidth}
\centering
Jongse Park\\
\textit{KAIST}\\
Daejeon, Republic of Korea\\
\href{mailto:jspark@casys.kaist.ac.kr}{\textcolor{blue}{jspark@casys.kaist.ac.kr}}
\end{minipage}
\end{tabular}
}

\begin{document}
\maketitle

\begingroup
\renewcommand{\thefootnote}{}
\footnotetext{\textsuperscript{\textdagger}These authors contributed equally.}
\endgroup

\begin{abstract}

System-level simulation is an essential tool for exploring the rapidly
expanding design space of LLM serving systems, where real deployments
remain costly and often infeasible.
However, modern LLM serving now evolves faster than human-driven
simulator development can track, and emerging workloads and
mechanisms, from agentic workflows to disaggregated serving, no longer
fit the monolithic simulation pipeline that existing simulators
assume.
Each new mechanism therefore demands an invasive rewrite, leaving a
widening development gap between deployed serving systems and the
simulators that model them.

To close this gap, we present \sysname{}, a framework that realizes
agent-driven simulator development.
\sysname{} introduces a \emph{composable simulator infrastructure}
that uniformly expresses the complete serving workflow, including the
control decisions that coordinate it, and realizes it as a
unified dynamic graph in \substrate{}.
\synthesizer{}, a harnessed coding agent, then lowers natural-language
feature requests onto this abstraction under simulator-specific
guardrails and fidelity validation, evolving one shared simulator
instead of building a new one for every feature.
Under the same coding agent and harnesses, extensions built on
\sysname{} follow a vLLM-based real system with 2.51\% average
throughput error, versus 6.03\% for extensions built on existing
simulators.
On identical workloads, \sysname{} also simulates up to
284.96$\times$ and 23.19$\times$ faster than two state-of-the-art
simulators, LLMServingSim2.0 and Vidur, respectively.

\end{abstract}

\section{Introduction}
\label{sec:intro}

Designing efficient LLM serving systems requires exploring an expanding design space spanning hardware, scheduling, and system optimizations. 
Evaluating these choices on real systems is often prohibitively expensive and difficult to scale, especially as emerging hardware accelerators are not yet available as off-the-shelf platforms. 
Even when deployable, bringing such systems up requires substantial engineering effort and cost. 
As a result, system-level simulation has become an essential tool for LLM serving research, enabling rapid and cost-effective exploration of design trade-offs~\cite{llmservingsim,llmservingsim2,vidur}.

\begin{figure}[!t]
    \centering
    \includegraphics[width=0.9\columnwidth]{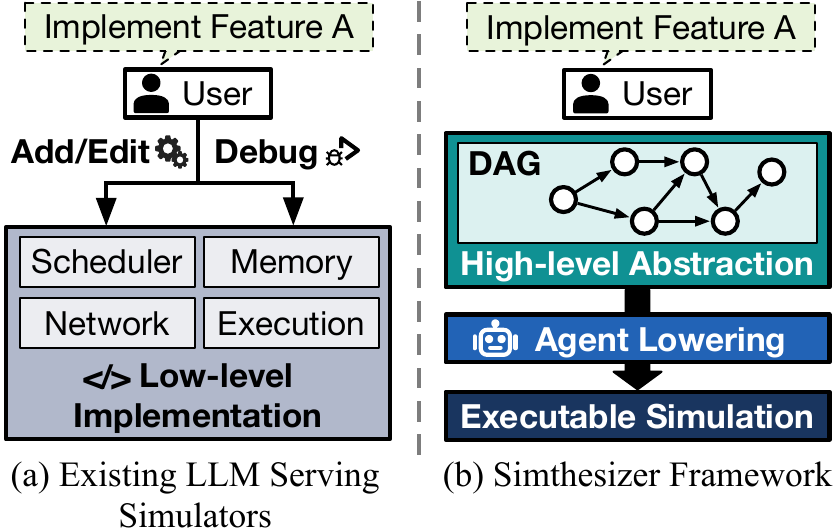}
    \caption{
        (a) Existing LLM serving simulators require manually implementing new serving mechanisms as they emerge, leading to low-level modifications, while (b) \sysname{} framework instead composes the complete serving workflow from uniform elements of a unified dynamic DAG, which is automatically lowered into executable simulation through agent-driven development.
    }
    \vspace{-2ex}
    \label{fig:intro}
\end{figure}

However, these simulators are now challenged by a fundamental shift in LLM serving, from the traditional single-inference execution flow to diverse and dynamic workflows.
This shift transforms both the pace at which serving systems evolve and the way requests and serving mechanisms are structured, demanding a rethinking of system-level simulator design.
Specifically, it introduces two key changes:

\begin{enumerate}[labelindent=0.3em,nolistsep,leftmargin=1.0em]
    \item \textbf{Fast evolution:}
    The LLM ecosystem evolves at a fast pace as new models, applications, and system-level optimizations continuously emerge.
    Serving techniques and execution patterns consequently change over time, rapidly invalidating a simulator's assumptions about system behavior and requiring repeated manual effort to keep it up to date.

    \item \textbf{Non-monolithic serving:}
    Agentic requests expand a single query into multiple model invocations interleaved with external tool calls and decision-making stages, following dynamic execution paths determined by intermediate results.
    Meanwhile, serving mechanisms such as disaggregated execution and speculative decoding break the once-monolithic inference flow into multiple interacting stages.
    Both trends replace the predetermined serving flow that existing simulators assume with dynamically composed stages.

\end{enumerate}

To cope with these challenges, it is natural to ask whether existing LLM serving simulators can keep up, but they are fundamentally limited.
Most simulators rest on two premises, that serving systems evolve slowly enough for manual engineering to keep pace and that serving behavior can be captured by a single monolithic simulation pipeline.
As the two changes above invalidate these premises, every non-monolithic mechanism now forces an invasive restructuring of the fixed simulation pipeline, while tracking fast evolution leaves developers repeatedly modifying and validating simulator internals.

To address these limitations, we propose a novel framework, \sysname, for developing modern LLM serving simulators, as illustrated in Figure~\ref{fig:intro}.
Our approach captures non-monolithic serving through a \emph{composable and extensible simulator infrastructure} that uniformly expresses the complete serving workflow, and tracks fast evolution through \emph{agent-driven lowering} that translates natural language specifications into executable simulation.
Our contributions are as follows:

\begin{description}[leftmargin=1.5em]

\item \textbf{(1) Composable and extensible simulator infrastructure for LLM serving simulation.}
We introduce \substrate{}, a simulator infrastructure that expresses every element of the serving workflow, rather than only its compute and communication operations, in one uniform, composable form.
\substrate{} realizes this design as a directed acyclic graph (DAG) where logical nodes express control decisions such as request routing and batch formation alongside compute and communication nodes.
A modular control layer implements these decisions as interchangeable components, so new serving mechanisms are integrated by composing components and reconnecting nodes without restructuring the execution engine.

\item \textbf{(2) Agent-driven lowering for scalable simulator extension.}
We develop \synthesizer{}, a harnessed coding agent that systematically lowers high-level specifications into executable simulation logic, bridging the gap between specified behavior and low-level execution.
Instead of manually implementing new system behaviors, our approach exposes structured interfaces and modular components that agents extend or compose, guided by harness engineering that constrains task scope, enabling reliable integration of emerging serving mechanisms.

\item \textbf{(3) End-to-end implementation and verification of an extensible simulator.}
We implement \sysname{} end-to-end, combining \substrate{} and \synthesizer{} into a complete framework for developing and extending LLM serving simulators.
To extend the simulator, \synthesizer{} refines a natural language feature request into a specification, maps it onto \substrate{}'s abstractions, implements any missing functionality, and validates the result against real-system measurements or reference evidence.
We demonstrate this workflow by extending \substrate{} with \synthesizer{} across modern serving mechanisms, including KV cache quantization, speculative decoding, and hybrid Mamba model support, evolving one shared simulator rather than building a new one per feature.

\end{description}

We instruct OpenAI Codex~\cite{codex}
to serve as \synthesizer{}, extending \substrate{} with new serving features and implementing the same features on existing simulators with the same set of harnesses.
Comparing the resulting extensions, we find that \sysname{} more closely follows the behaviors of a vLLM-based real system, achieving an average throughput error of 2.51\%, whereas extensions built on existing simulators exhibit a significantly higher throughput error of 6.03\%.
Lastly, using identical LLM workloads, \sysname{} achieves up to 284.96$\times$ and 23.19$\times$ faster simulation time than two state-of-the-art simulators, LLMServingSim2.0~\cite{llmservingsim2} and Vidur~\cite{vidur}, respectively.

These results demonstrate that our framework effectively supports the development and extension of modern LLM serving simulators, enabling accurate and extensible modeling of complex and rapidly evolving workloads.
While our evaluation focuses on LLM serving simulation, the combination of composable simulator infrastructure and agent-driven lowering suggests a promising direction for building simulators in other systems domains.
We view this work as an initial step toward such a direction, and hope it encourages further exploration of more scalable and automated simulation methodologies.
\sysname{} is available at \href{https://github.com/casys-kaist/Simthesizer}{\textcolor{blue}{https://github.com/casys-kaist/Simthesizer}}.

\section{Background and Motivation}
\label{sec:background}

\subsection{LLM Serving Simulation}
\label{subsec:bg:simulators}

\niparagraph{Challenges in evaluating LLM serving systems.}
As the scale and complexity of LLM serving systems continue to grow,
their design space has expanded to encompass a wide range of
deployment choices, including parallelism
strategies~\cite{dist_serve, splitwise, loop_serve}, scheduling
policies~\cite{orca, sarathi-serve, llumnix}, batch
formation~\cite{orca, sarathi-serve, nanoflow}, KV-cache
management~\cite{vllm, cached_attention, asplos25_pod_attention,
vattention}, and workload composition~\cite{pensieve, deltazip}.
However, evaluating these choices on real systems is often
prohibitively expensive and difficult to scale.
Moreover, many candidate designs, including emerging accelerators,
interconnects, and cluster configurations, are not yet available for
direct measurement.

\niparagraph{Simulation for LLM serving.}
To address these challenges, the systems community has turned to
system-level simulators as a practical means of exploring design
trade-offs in LLM serving systems~\cite{vidur, apex, llmservingsim,
llmservingsim2}.
These simulators allow researchers to explore and validate early ideas across scheduling, hardware architectures, and system configurations before implementation or deployment~\cite{splitwise, pascal, dist_serve, pimba}.

Vidur~\cite{vidur} predicts LLM inference performance across
parallelism strategies, batch sizes, and scheduling policies using a
random-forest model trained on profiled GPU operator latencies.
Similarly, APEX~\cite{apex} predicts performance across models,
quantization formats, batching policies, and device-cluster
configurations.
To do so, it constructs candidate serving configurations using
predefined modules and feeds them to its simulator.
LLMServingSim~\cite{llmservingsim} and
LLMServingSim2.0~\cite{llmservingsim2} model additional modern serving
mechanisms, including prefix caching, prefill/decode disaggregation, and
KV-cache offloading, through higher-level abstractions.

\niparagraph{Two outdated premises.}
Although these simulators adopt different abstractions, they all encode serving behavior as components that human developers implement in advance and integrate into a fixed simulation pipeline.
This shared design remains practical only under two premises: (1) the
target serving system evolves \emph{slowly}, and (2) serving behavior
can be captured by a \emph{monolithic} simulation pipeline.
The first premise assumes that new serving behaviors arise infrequently
enough to amortize the substantial human effort required
to implement them.
The second assumes that a single, tightly integrated control flow can
encode diverse serving techniques as predefined policies.
As we discuss in Section~\ref{subsec:bg:shift}, however, the
rapid evolution of serving mechanisms and the shift toward
non-monolithic serving call both premises into question.

\subsection{The Shifting Landscape of LLM Serving}
\label{subsec:bg:shift}

Modern LLM serving is shifting along two coupled directions: (1) its
workloads and mechanisms evolve at an unprecedented pace, and (2) many
of them no longer take the form of a monolithic inference flow.

\begin{figure}[t]
    \centering
    \includegraphics[width=0.85\linewidth]{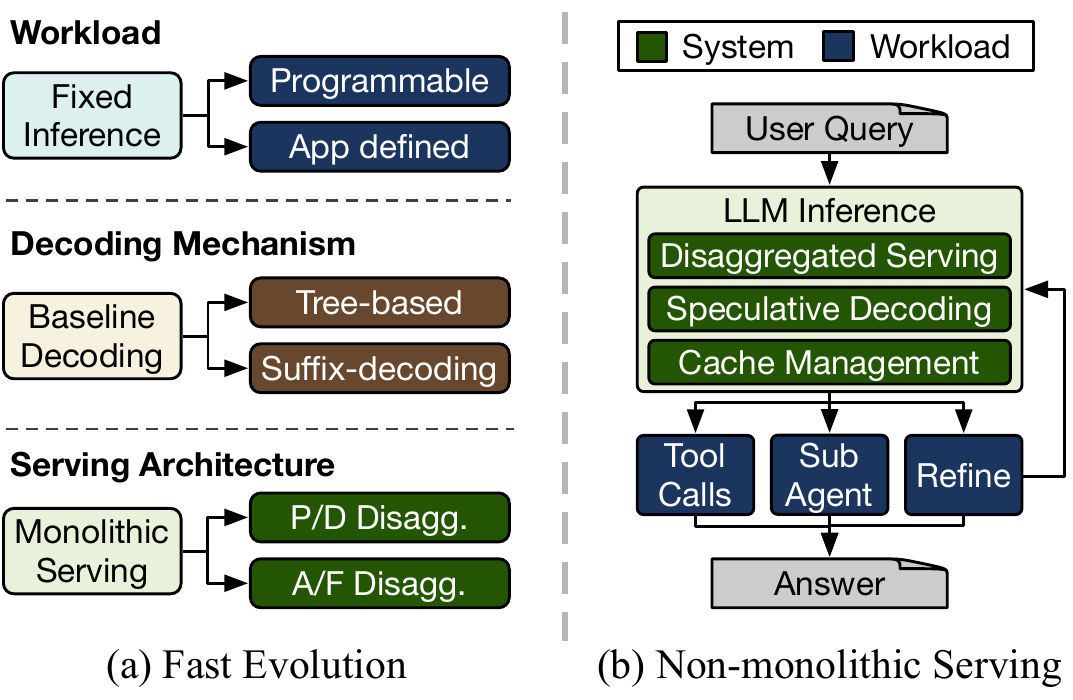}
    \caption{Shifts in modern LLM serving: (a) fast evolution and (b)
    non-monolithic serving.}
    \vspace{-3ex}
    \label{fig:agent-vs-llm}
\end{figure}

\begin{figure*}[t]
  \centering
  \includegraphics[width=0.9\textwidth]{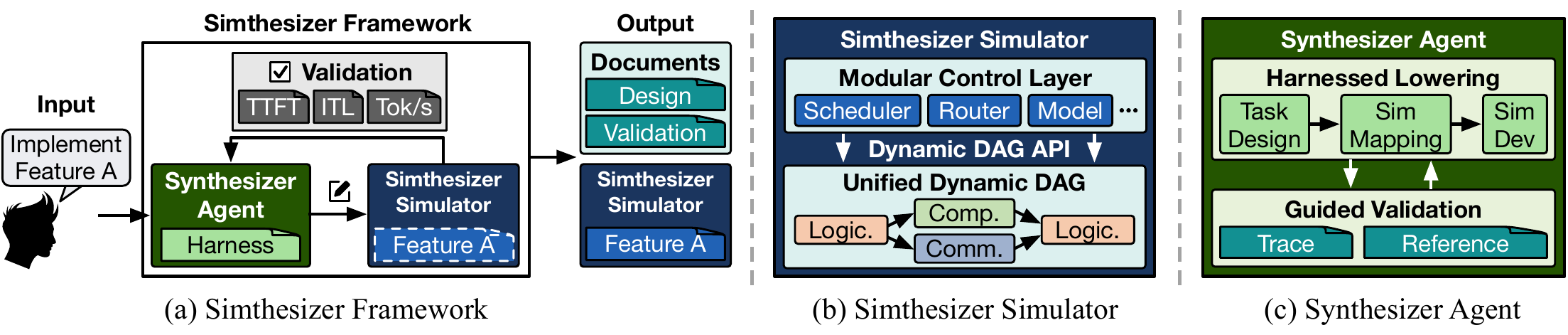}
  \vspace{-1ex}
  \caption{Overview of the \sysname{} framework.}
  \vspace{-2ex}
  \label{fig:sim_overview}
\end{figure*}

\niparagraph{Fast evolution.}
First, LLM workloads and serving mechanisms now evolve rapidly and
concurrently across every level of the serving stack.
In just a few years, agentic workloads have pushed serving systems
from fixed inference flows~\cite{vllm} toward programmable,
application-defined workflows~\cite{parrot, pie}.
Over the same period, speculative decoding has raced through
successive generations, from conventional draft-and-verify methods to
tree-based and suffix-decoding schemes~\cite{spec_dec, specinfer,
eagle1, eagle2, eagle3, suffix_decoding}.
Serving architectures likewise keep diversifying toward P/D and
attention--FFN disaggregation~\cite{dist_serve, splitwise,
megascale_infer}.
Each new wave arrives before the previous one settles into standard
practice.

\niparagraph{Non-monolithic serving.}
Second, many emerging workloads and serving mechanisms no longer fit
within a single, self-contained inference lifecycle.
On the workload side, as shown in Figure~\ref{fig:agent-vs-llm},
agentic requests expand a single user query into multiple LLM
inferences interleaved with external tool calls, sub-agent execution,
and iterative context refinement~\cite{react, toolformer, hugging_gpt,
reflexion, autogen, metagpt}.
A request is thus no longer a single inference call but a
runtime-dependent composition of inference and non-inference stages.
On the system side, disaggregated serving splits prefill and decode
across separate machine pools~\cite{dist_serve, splitwise},
speculative decoding coordinates draft and target models within a
decoding loop~\cite{spec_dec, specinfer}, and KV-cache management
distributes request state across turns and storage
tiers~\cite{flashgen, kvcache_cache, cached_attention}.
Despite optimizing different resources, they all replace the
self-contained inference lifecycle with orchestration across
separately managed components.

\niparagraph{Implications for existing simulators.}
These two trends undermine the premises of existing simulators.
New serving behaviors emerge faster than human developers can
implement, integrate, and validate them, leaving a widening
development gap between deployed serving systems and the
simulators that model them.
Meanwhile, non-monolithic serving breaks the fixed serving loop that
existing simulators assume every request follows
(Section~\ref{subsec:bg:simulators}).
A feature that falls outside this loop therefore requires
restructuring its control flow and propagating changes across
existing paths.

\subsection{Rethinking Simulation Design for Modern LLM Serving}
\label{subsec:bg:requirements}

\niparagraph{Agent-driven simulator development.}
Both implications ultimately stem from a shortage of engineering
labor. %
Fast evolution multiplies how often developers must extend a
simulator to close the development gap, while non-monolithic serving
turns each extension into an invasive rewrite.
Sustaining simulators against both pressures calls for automating the
implementation work itself. %
Recent LLM-based coding agents make this automation practical, and
the systems community already employs them~\cite{aiopslab, looprag, knighter}.
We therefore envision \emph{agent-driven simulator development}, in
which coding agents implement new serving mechanisms as they emerge.

However, simply adding an AI agent to existing simulators is
insufficient.
Bound to a monolithic pipeline, an agent inherits the same structural
coupling, so each extension remains a broad, system-wide rewriting that is difficult to review and may introduce silent regressions.
Keeping modifications local requires a representation that captures inference and non-inference stages, and the control flow coordinating them within a single abstraction.
Even then, generated code that compiles and passes functional tests
can still model the wrong system, demanding simulator-specific
guardrails and fidelity validation.
From these requirements, we derive two design principles for
next-generation serving simulators:
\begin{itemize}[labelindent=0.3em,nolistsep,leftmargin=1.0em]
    \item \textbf{(Principle 1) Composable simulation from a uniform
        representation.}
        Existing simulators bind serving mechanisms to technique-specific
        control paths, making each extension a system-wide change.
        A modern simulator should instead express workloads and
        mechanisms in a uniform, composable representation that
        captures stages, dependencies, and state transitions, allowing
        humans or agents to add and compose components without
        restructuring the execution engine.
    \item \textbf{(Principle 2) Guarded agent-driven implementation.}
        Structural locality bounds an agent's changes, but does not
        ensure simulation fidelity, as an extension can compile and
        produce plausible outputs while omitting performance-critical
        state or violating simulator-specific invariants.
        A modern simulator must therefore pair bounded extension
        interfaces and simulator-specific guardrails with validation
        methods that evaluate simulation semantics.
\end{itemize}

\section{Overview of \sysname{}}
\label{sec:overview}

Guided by these principles, we present \sysname{}, a simulation
framework that realizes agent-driven simulator development for modern
LLM serving.
\sysname{} pairs a composable, extensible simulation substrate
(Principle~1) with a guarded agent workflow (Principle~2), as
illustrated in Figure~\ref{fig:sim_overview}.

\niparagraph{\sysname{} framework.}
\sysname{} comprises two components, (1) \substrate{}, a unified and
extensible LLM serving simulator, and (2) \synthesizer{}, a harnessed
coding agent that integrates new serving mechanisms into \substrate{}.
The workflow begins with a user request specifying the feature to
implement.
\synthesizer{} refines the intended behavior, maps it onto
\substrate{}'s abstractions, and implements any functionality
\substrate{} does not already provide.
Then, \sysname{} validates the observed behavior against real-system measurements or reference evidence.
The workflow finally returns the extended \substrate{} together with
inspectable artifacts exposing its modeling decisions,
implementation decisions, and validation results.

\niparagraph{\substrate{}.}
Existing simulators use static DAGs and pipelines to encode compute and
communication operations, so any mechanism that alters this pipeline demands an invasive, end-to-end redesign.
To address this, \substrate{} composes the entire serving workflow
from a uniform set of elements, using compute, communication, and
logical nodes of a unified dynamic DAG to capture both LLM operations
and the control decisions. %
Through Dynamic DAG APIs, logical nodes insert operations and
rewire dependencies at runtime, allowing \substrate{} to integrate new
mechanisms and express runtime-dependent workflows without modifying the execution engine.
Above this substrate, a modular control layer implements serving
policies through interchangeable components such as schedulers,
routers, model runners, and resource simulators.
Existing components compose through structured configuration, and a
new mechanism requires implementing only the behavior the control
layer lacks (Section~\ref{sec:design}).

\niparagraph{\synthesizer{}.}
\synthesizer{} is a coding agent with a harness tailored to
developing LLM serving simulators.
It lowers a user request through three stages, \texttt{task-design},
\texttt{sim-mapping}, and \texttt{sim-dev}, which respectively produce
a simulator-facing specification, map each decision to a \substrate{}
abstraction, and extend the implementation.
The agent then evaluates simulation fidelity through trace-guided
validation when real-system measurements are available and
reference-guided validation otherwise.
Users may resolve performance-critical ambiguities and review
the specification, implementation map, and validation report, while
\synthesizer{} performs the low-level implementation and revision
work.
Together, \substrate{} and \synthesizer{} make simulator extension
both compositional and reviewable, allowing \sysname{} to track new
serving mechanisms without restructuring a monolithic simulation
pipeline (Section~\ref{sec:harness-v3}).

\section{\sysname{} Simulator}
\label{sec:design}

\begin{table*}[t]
  \caption{Dynamic DAG modification APIs.}
  \centering
  \footnotesize
  \begin{tabular}{p{0.5\linewidth}p{0.45\linewidth}}
    \toprule[1.2pt]
    \textbf{Interface} & \textbf{Description} \\
    \midrule[1.2pt]
    \parbox[t]{\linewidth}{%
      {\ttfamily\bfseries add\_logical\_node}{\ttfamily(operation,
      state) -> node} \\
      {\ttfamily\bfseries add\_logical\_node\_at}{\ttfamily(operation,
      state, at) -> node}%
    } &
    Insert a logical node carrying a policy operation and its state.
    The \texttt{\_at} variant seeds an explicit start time, e.g., for
    request arrival. \\
    \midrule
    {\ttfamily\bfseries add\_compute\_node}{\ttfamily(layer, batch)
    -> node} &
    Insert a compute node tagged with a semantic layer descriptor and
    the scheduled batch state. \\
    \midrule
    {\ttfamily\bfseries add\_network\_node}{\ttfamily(src, dst, bytes)
    -> node} &
    Insert a point-to-point transfer between two network devices.
    Its timing model resolves latency, including multi-hop routing when
    configured. \\
    \midrule
    {\ttfamily\bfseries add\_edge}{\ttfamily(parent, child)} &
    Declare a precedence constraint.
    A child becomes runnable only after all of its predecessors complete. \\
    \midrule
    {\ttfamily\bfseries pop\_edges}{\ttfamily(node) -> list[node]} &
    Detach the current node's outgoing edges so the caller can splice
    a new subgraph before the original successor. \\
    \bottomrule[1.2pt]
  \end{tabular}

  \label{tab:event-substrate-api}
  \vspace{-1.5em}
\end{table*}

\begin{figure}[t]
  \centering
  \includegraphics[width=0.95\linewidth]{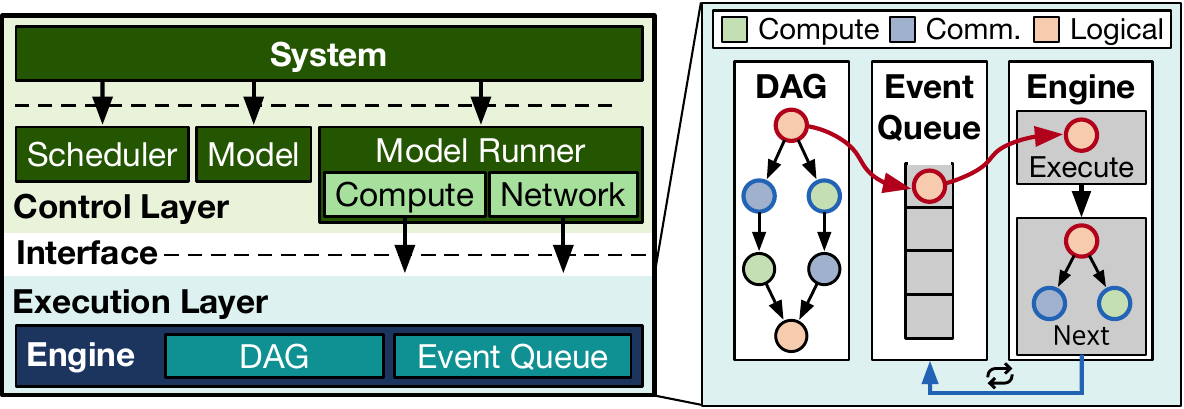}
  \caption{Architecture of \substrate{}.}
  \vspace{-3ex}
  \label{fig:foundation_simulator}
\end{figure}

Figure~\ref{fig:foundation_simulator} shows the two-layer architecture of the \substrate{}.
The execution layer realizes the composable simulator infrastructure
as a unified dynamic DAG that represents an LLM serving
workflow, allowing new mechanisms to extend the graph without
restructuring the execution engine.
The control layer constructs and modifies these graphs through modular
components that implement serving policies.
This separation gives \substrate{} a stable execution substrate while
its serving behavior evolves through localized extensions.

\subsection{Unified Dynamic DAG Abstraction}
\label{subsec:design:abstractions}

\niparagraph{Execution graph representation.}
\substrate{} represents the complete simulated execution as a unified
dynamic DAG with three node types: (1) \emph{compute nodes} that
specify which model operation executes on which device, (2)
\emph{communication nodes} for inter-device transfers, and (3)
\emph{logical nodes} for policy events such as request arrival and
scheduling.
Each edge encodes an execution-order dependency, and simulation
executes each node once all of its predecessors complete.

\niparagraph{Node taxonomy.}
At execution, each node differs along two dimensions: (1) latency is
\emph{deterministic} when fixed at dispatch and
\emph{non-deterministic} when later activity can change the completion
time, and (2) behavior is \emph{stateful} when execution alters
subsequent control flow or updates serving state and \emph{stateless}
otherwise.
The three node types occupy distinct points in this taxonomy, which
determines how the execution layer processes each of them.

\niparagraph{Compute nodes.}
\substrate{} treats compute nodes as deterministic and stateless by
construction.
A compute node represents an already scheduled model operation on a
target device, with shared-device ordering encoded by its dependencies.
Once runnable, its operation, batch, and device model determine a
single latency, and kernel-level effects remain encapsulated within
that latency model.

\niparagraph{Communication nodes.}
Communication nodes are non-deterministic and stateless, as link
contention can affect their completion time while their execution
leaves state unchanged.
To model non-deterministic latency, \substrate{} allows the network
model to invalidate a node's scheduled completion event when resource
state changes and to provide a revised completion time for
rescheduling.
This event-invalidation interface supports dynamic contention without
binding the execution layer to a particular network model.

\niparagraph{Logical nodes.}
Logical nodes are deterministic and stateful.
They encode control decisions such as request routing, batch
formation, and state transitions, with a fixed execution latency and
explicit updates to request or system state.
Existing simulators implement these decisions in fixed control flow
outside their compute and communication DAGs; \substrate{} instead
makes them first-class nodes in the execution graph.
A new serving workflow can therefore be constructed by adding and
reconnecting compute, communication, and logical nodes, leaving the
core execution engine untouched.

\niparagraph{Event-driven processing.}
The execution layer processes all node types through a common
event-driven loop.
It first initializes simulated time and inserts every node with no
unresolved predecessor into a time-ordered event queue.
At each step, it dequeues the earliest event, advances simulated time,
and processes the corresponding node.
A deterministic node schedules a fixed completion event at dispatch,
whereas a change in resource state can invalidate and replace the
completion event of a non-deterministic node.
A stateful node may additionally update serving state or modify the
execution graph.
Upon completion, the execution layer propagates the node's completion
time to its successors, resolves incoming dependencies, and
enqueues each newly runnable node.

\subsection{Dynamic DAG APIs}
\label{subsec:design:dag}

LLM serving is inherently dynamic, as batch composition, placement,
and even the stages a request traverses depend on runtime state, so a
simulator cannot enumerate every operation and dependency in advance.
Existing simulators sidestep this dynamism by keeping runtime
decisions in a fixed simulation pipeline outside their compute and
communication DAGs~\cite{llmservingsim2, llmservingsim, vidur, apex}.
\substrate{} instead lets logical nodes grow the graph as decisions
become known, using the Dynamic DAG APIs in
Table~\ref{tab:event-substrate-api} to insert new nodes and rewire
dependencies, keeping workflow changes localized.

\niparagraph{Node construction.}
Dynamic expansion first materializes each runtime-selected action as
one of the three node types.
\texttt{add\_logical\_node} records the operation and state for a
policy event, \texttt{add\_compute\_node} records the semantic layer
and scheduled batch, and \texttt{add\_network\_node} records the
source, destination, and transfer size.
Each API returns a node handle for later connections, and these typed
constructors keep newly added nodes consistent with the execution
semantics of the initial graph.

\niparagraph{Dependency construction.}
Creating nodes alone does not determine when they execute; the
\texttt{add\_edge(parent, child)} API specifies that the child can run
only after the parent completes.
A logical node may also alter existing dependencies; in disaggregated
serving, for example, a decode stage already follows its prefill
stage, but assigning decode to another machine requires a KV-cache
transfer to complete in between.
The \texttt{pop\_edges} API detaches and returns the prefill node's
outgoing edges, letting the logical node splice in the
transfer node and reconnect the decode stage after it.
This local rewrite preserves the surrounding graph while enforcing the
new execution order.

\subsection{Modular Control Layer}
\label{subsec:design:control}
\label{subsec:design:ownership}

\niparagraph{Components own serving roles.}
The Dynamic DAG APIs define how the execution graph can grow at
runtime, but not who grows it.
A single handler making every serving decision would recreate a
monolithic pipeline above the execution layer, forcing each new
mechanism to modify shared control code.
To address this, \substrate{} partitions the control layer into
components, each owning one serving role such as request routing,
batch scheduling, or model execution.
Each component keeps its role's policy state, handles that role's
logical nodes, and materializes its decisions through the Dynamic DAG
APIs.
This ownership confines each policy change to the component that
makes the decision.

\begin{figure}[t]
  \centering
  \begin{minipage}{\linewidth}
\begin{examplecode}
interface System extends Logical:
  into_results() -> list[result]
  Logical::handle(...) -> time

interface Scheduler:
  enqueue_sub_request(sub_request)
  schedule() -> batch
\end{examplecode}
  \end{minipage}
  {\small (a) Component interfaces\par}
  \begin{minipage}{\linewidth}
\begin{examplecode}
impl System for SingleInstance:
  scheduler: Scheduler

  def into_results() -> list[result]
  def Logical::handle(...) -> time

impl Scheduler for ChunkedPrefill:
  enable_prefix_caching: bool

  def enqueue_sub_request(sub_request)
  def schedule() -> batch
\end{examplecode}
  \end{minipage}
  {\small (b) Component implementations\par}
  \begin{minipage}{\linewidth}
\begin{examplecode}
system:
  kind = "SingleInstance"
  scheduler:
    kind = "ChunkedPrefill"
    enable_prefix_caching = true
\end{examplecode}
  \end{minipage}
  {\small (c) Compositional configuration\par}
  \caption{Examples of (a) abstract interfaces capturing general serving behaviors; (b) concrete implementations of interfaces with encapsulated states; and (c) structured configuration composing implementations and defining component-specific parameters.}
  \label{fig:modular_interfaces}
\end{figure}

\niparagraph{Interfaces capture stable roles.}
Serving roles remain stable even as the policies filling them change
rapidly; every serving system forms batches, but the batching policy
evolves with each new mechanism.
\substrate{} therefore fixes each role's operations in an interface
and leaves policy and state to concrete implementations.
In Figure~\ref{fig:modular_interfaces}(a), \texttt{System}
orchestrates one serving instance, handling its logical events
through \texttt{handle} and emitting results through
\texttt{into\_results}.
\texttt{Scheduler} owns iteration-level batching, receiving
sub-requests through \texttt{enqueue\_sub\_request} and returning the
next batch through \texttt{schedule}.
Because \texttt{System} depends only on this interface, any
conforming scheduler can fill the same position in the hierarchy.

\niparagraph{Implementations encapsulate policy state.}
A policy is more than a function; it carries state such as request
queues, cache metadata, and tuning parameters.
If this state leaked across components, replacing one policy would
ripple through the others.
\substrate{} therefore confines each policy's state to the
implementation that realizes it
(Figure~\ref{fig:modular_interfaces}(b)).
\texttt{ChunkedPrefill} implements \texttt{Scheduler} and privately
tracks its prefix-caching flag, while \texttt{SingleInstance}
implements \texttt{System}, holding its scheduler only through the
interface.
Swapping in another scheduler thus touches neither
\texttt{SingleInstance} nor the execution layer.

\niparagraph{Configuration composes a system.}
A structured configuration assembles these components into a complete
simulator (Figure~\ref{fig:modular_interfaces}(c)).
Each \texttt{kind} field selects an implementation for one interface,
nested entries select subcomponents, and the remaining fields set
implementation-specific parameters.
The example builds \texttt{SingleInstance} over \texttt{ChunkedPrefill}
and enables prefix caching without touching simulator code.
This composition bounds the cost of extending \substrate{}.
When every required policy exists, a user or \synthesizer{} composes
the target system through configuration alone; when one is missing,
\synthesizer{} implements only that policy behind its interface and
reuses the remaining components.
An extension is thus a configuration edit or a component-local
implementation, never a control-layer rewrite.

\subsection{Implementation}
\label{subsec:design:implementation}

\niparagraph{Built-in components.}
We instantiate the control layer with seven component interfaces that
mirror a request's path through a serving system, namely request
routing, system orchestration, batch scheduling, model execution,
model description, and compute and network timing.
This mirroring places every serving decision that a new mechanism may
change behind a distinct interface.
Behind these interfaces, \substrate{} ships built-in implementations
covering common serving configurations.

At the system level, built-ins cover single- and multi-instance
serving, prefill--decode disaggregation with explicit KV-cache
handoff, and expert-parallel MoE serving, with routing policies that
place requests round-robin or by per-instance load.
Along the execution path, the built-in scheduler implements chunked
prefill with memory-aware batch admission and block-aligned prefix
caching, model runners expand each scheduled batch into layer-level
compute and communication nodes under tensor or expert parallelism,
and model descriptions cover dense and MoE architectures.
These built-ins reduce common serving studies to configuration,
reserving \synthesizer{} for new mechanisms.

\niparagraph{Pluggable timing backends.}
The execution and control layers consume completion times without
depending on how they are produced.
This boundary lets detailed architecture and network
simulators~\cite{accel_sim, mgpusim, ns3, astra_sim} serve as backends
when their fidelity is required, while \substrate{} provides
lightweight defaults for end-to-end serving studies.

\niparagraph{Profile-based compute backend.}
The default compute backend estimates each compute node's latency from
offline profiles~\cite{llmservingsim, llmservingsim2, triosim, vTrain}.
Because a single total-token key cannot capture layers whose latency
depends on more than batch size, the backend indexes each layer type
with at most two layer-specific keys, such as the cached KV footprint
and query--context interaction count for attention or the token count
and activated experts for MoE.
The profiler samples these keys offline and interpolates over the table.

\niparagraph{Flow-based network backend.}
The default network backend models communication at flow granularity
rather than simulating individual packets~\cite{triosim}; each
transfer follows a routed path over a device--link connectivity graph
and receives the minimum available bandwidth along it.
When a flow starts or completes, the backend recomputes the affected
bandwidths and reschedules completion events through the
event-invalidation interface
(Section~\ref{subsec:design:abstractions}).
Because both backends sit behind component interfaces, either can be
replaced without changing the execution layer or serving
policies.

\section{Synthesizer Agent}
\label{sec:harness-v3}

 \begin{figure}[t]
    \centering
    \includegraphics[width=0.9\linewidth]{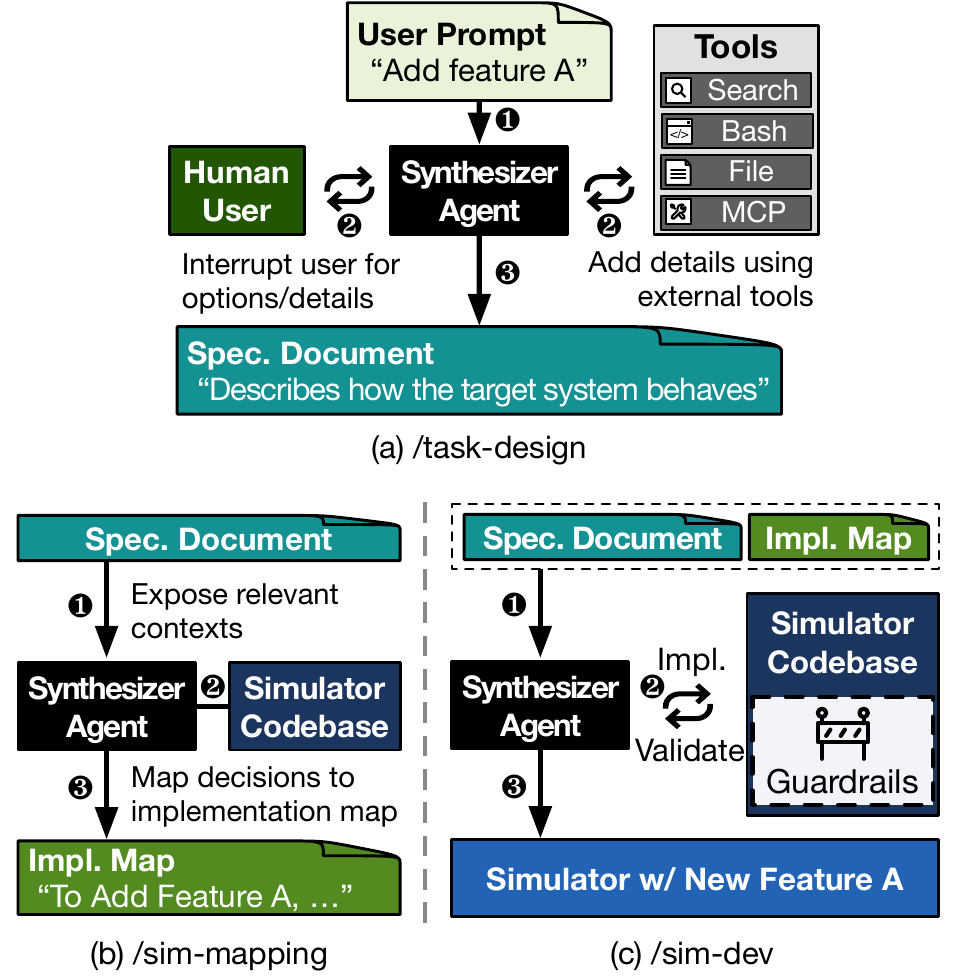}
    \caption{Agent-driven simulator extension process.}
    \vspace{-2ex}
    \label{fig:agent-harness}
\end{figure}

Grounded in \substrate{}, \sysname{} translates natural language
specifications into executable simulation through agent-driven
development that keeps pace with fast-evolving serving mechanisms.
\synthesizer{} realizes this lowering by wrapping a coding agent in a
harness tailored to developing LLM serving simulators, leveraging
\substrate{}'s modular interfaces to extend the simulator without
reworking its internals.
The harness makes simulator-specific constraints explicit and
organizes synthesis around three inspectable artifacts: a
specification, an implementation map, and a validation report.
This design shifts human judgment from low-level code editing to
resolving performance-critical ambiguities and reviewing
implementation decisions and validation evidence.
We first describe the synthesis requirements and lowering process,
then present two validation regimes and their human review points.

\subsection{Simulator-Specific Synthesis Requirements}
\label{subsec:harness-draft-en-contracts}
Synthesis begins when the user asks \synthesizer{} to integrate a new
serving mechanism.
However, the initial request alone rarely resolves the modeling
decisions needed for a faithful implementation, leaving
performance-critical choices implicit even though they directly
affect simulation behavior.
We identify three recurring requirements for translating the request
into an implementation that captures the intention: %

\begin{itemize}[labelindent=0.3em,nolistsep,leftmargin=1.0em]
    \item \textbf{Semantic completeness.}
        The simulator-facing specification must identify the
        quantities and assumptions that determine simulation time,
        resource contention, state lifetime, dependency structure,
        and metric boundaries.
        Each performance-relevant choice must be either resolved in
        the specification or surfaced explicitly for human review.

    \item \textbf{Structural alignment.}
        Each modeling decision must map to the \substrate{}
        abstraction responsible for the corresponding state
        transition, graph update, or resource interaction.
        This mapping avoids scattering a mechanism across unrelated
        control paths and preserves a traceable connection between
        the specification and the generated implementation.

    \item \textbf{Modeling fidelity.}
        The implementation must model the serving mechanism at the
        level of detail required by the target performance question.
        When exact modeling is unavailable, it may introduce
        approximations, such as aggregate multipliers or statistical proxies.
        The harness requires explicit assumptions, supporting
        evidence, and validity boundaries for each approximation.
\end{itemize}

\subsection{Harnessed Lowering Process}
\label{subsec:harness-draft-en-pipeline}
Figure~\ref{fig:agent-harness} shows \sysname{}'s lowering stages for
meeting the synthesis requirements: \texttt{task-design},
\texttt{sim-mapping}, and \texttt{sim-dev}.
Each stage addresses a distinct requirement and produces documents
that record the corresponding decisions.
This linkage routes each issue back to the stage where the relevant
decision was made, rather than treating every issue as an implementation bug.

\niparagraph{(1) \texttt{task-design}.}
The first stage targets semantic completeness by refining the initial
user request into a simulator-facing specification.
\synthesizer{} consults available mechanism descriptions, external
sources, and \substrate{} interfaces.
It specifies the target configuration, states, dependencies, metrics,
and the assumptions and evidence for any approximation.
The resulting specification document records resolved decisions and
explicit modeling boundaries and remains in the synthesis context
throughout subsequent stages.

\niparagraph{(2) \texttt{sim-mapping}.}
The second stage targets structural alignment by mapping the
simulator-facing specification to the corresponding \substrate{}
components and interfaces.
The resulting implementation map records where state resides, which
Dynamic DAG operations encode dependencies, which components model
compute and communication, and which observable signals support
subsequent validation.
Rather than a sequence of code edits, the map ties each modeling
decision to the \substrate{} abstraction responsible for the
behavior, its implementation site, and its validation strategy.

\niparagraph{(3) \texttt{sim-dev}.}
The third stage implements the changes described in the
implementation map and checks for deviations.
\synthesizer{} edits the simulator components and iterates on
compiler diagnostics, runtime failures, test results, and violations
of the specification.
Simulator-specific guardrails detect undeclared approximations,
unexplained constants, and code changes that depart from the implementation map.
This stage records the implemented behavior, approximations, and
uncertainties in the validation report for human review.

\niparagraph{Synthesis-time human involvement.}
Most revisions proceed without continuous human intervention.
Incorrect component mappings rewind the lowering process to
\texttt{sim-mapping}, while implementation bugs are resolved within
\texttt{sim-dev}.
During \texttt{task-design}, the available references and simulator
context may be insufficient to resolve a performance-critical modeling choice.
In that case, \synthesizer{} requests human clarification of the
intended behavior, missing parameters, and modeling boundaries rather
than silently selecting a default.
At the end of the synthesis run, the user reviews the generated
documents and, if necessary, revises the specification and initiates
another run.

\begin{figure}[t]
    \centering
    \begin{minipage}{\linewidth}
\begin{specdoc}
[Target System Config]
scheduler           = chunked_prefill + prefix_caching
speculative_width   = $K=2$

[Modeling Input]
algorithm_reference = EAGLE-3
acceptance_model    = empirical distribution
sampling            = stochastic

[Simulator Semantics]
remaining_output    = $R$
verification_width  = $W=\min(K+1,R)$
accepted_drafts     = $A\sim p_{\mathrm{acc}}$
committed_progress  = $M=A+1$
target_compute      = evaluate $W$ positions
persistent_state    = advance and retain $M$ tokens

[Validation Protocol]
min_validation_step = 3
validation_data     = ShareGPT trace, system data
\end{specdoc}
    \end{minipage}
    \caption{Structured summary of the document produced by
    \texttt{task-design}. The full specification presents the same content in
    natural language.}
    \vspace{-2ex}
    \label{fig:case-study-specification}
\end{figure}

\subsection{Evidence-Guided Validation}
\label{subsec:harness-draft-en-validation}
The harnessed \synthesizer{} produces a candidate simulator that is
consistent with its specification and implementation map.
However, this consistency does not guarantee that the design itself
faithfully captures the target mechanism.
\sysname{} addresses this gap through evidence-guided validation,
which comprises the following two complementary regimes.

\niparagraph{Trace-guided validation.}
When real-system measurements are available, \sysname{} compares the
candidate simulator with the system under the same workload, serving
configurations, and hardware settings.
It aligns internal execution signals, such as queue length, batch
composition, and resource utilization, to identify the modeling
decisions responsible for the observed discrepancies.
Together with the implementation map, these signals help associate
each divergence with the corresponding scheduling, compute,
communication, or dependency behavior.
The validation report records the observed discrepancies, their
possible sources, and the modeling decisions that require revision,
providing concrete targets for the next synthesis run.

\niparagraph{Reference-guided validation.}
When comparable measurements are unavailable, as in cases involving
unreleased hardware, unavailable software stacks, or novel
mechanisms, \sysname{} performs reference-guided validation by
grounding its modeling decisions in available technical evidence.
\synthesizer{} derives performance criteria from reference
implementations, results reported in prior works, and other evidence,
then assesses whether the candidate simulator reproduces the expected
behavior under comparable conditions.
The validation report identifies evidence-grounded decisions,
unresolved assumptions, and approximations that define the validity boundary.
Although this regime does not establish empirical agreement with the
target system, it provides traceable technical justification and
exposes the remaining uncertainty.

\section{Simulator Synthesis in Action}
\label{sec:foundation-guided-synthesis}

We explain how \sysname{}'s two core components jointly support a
concrete simulator extension using EAGLE-style speculative
decoding~\cite{eagle3} as a case study.
This mechanism uses a lightweight drafter model to rapidly generate
several draft tokens.
The target model then verifies drafted tokens in a single inference
iteration, while only the accepted tokens advance the generation.
Although this mechanism is concise at the algorithmic level, its
simulation must distinguish the work performed by the target model
from the progress made by each request.
We trace the interaction of \substrate{} and \synthesizer{} from the
initial user request through semantic refinement, lowering, and
validation-driven revision.
Across five synthesis runs, we follow a representative trial through
specification, mapping, implementation, and validation.
\begin{table}[t]
    \centering
    \caption{Implementation map for speculative decoding.}
    \label{tab:spec-mapping}
    \footnotesize
    \setlength{\tabcolsep}{5pt}
    \renewcommand{\arraystretch}{1.16}
    \begin{tabularx}{\columnwidth}{
            @{}
            >{\raggedright\arraybackslash\bfseries}p{0.27\columnwidth}
            >{\raggedright\arraybackslash}X
            @{}
        }
        \toprule[1.2pt]
        \textbf{Behavior} & \textbf{Implementation map} \\
        \midrule[1.2pt]

        Target model verification &
        Route \(W\) through the existing request representation to
        the model runner and compute simulator. \\

        \midrule

        Committed request state &
        Assign \(M\) to scheduler-managed request and KV state,
        exposing only the committed prefix to the existing cache interface. \\

        \midrule

        Speculation round execution &
        Reuse the common DAG scheduling and execution path without
        adding a mechanism-specific event loop. \\

        \bottomrule[1.2pt]
    \end{tabularx}
    \vspace{-1.5em}
\end{table}

\begin{figure*}[t]
    \centering
    \includegraphics[width=0.9\linewidth]{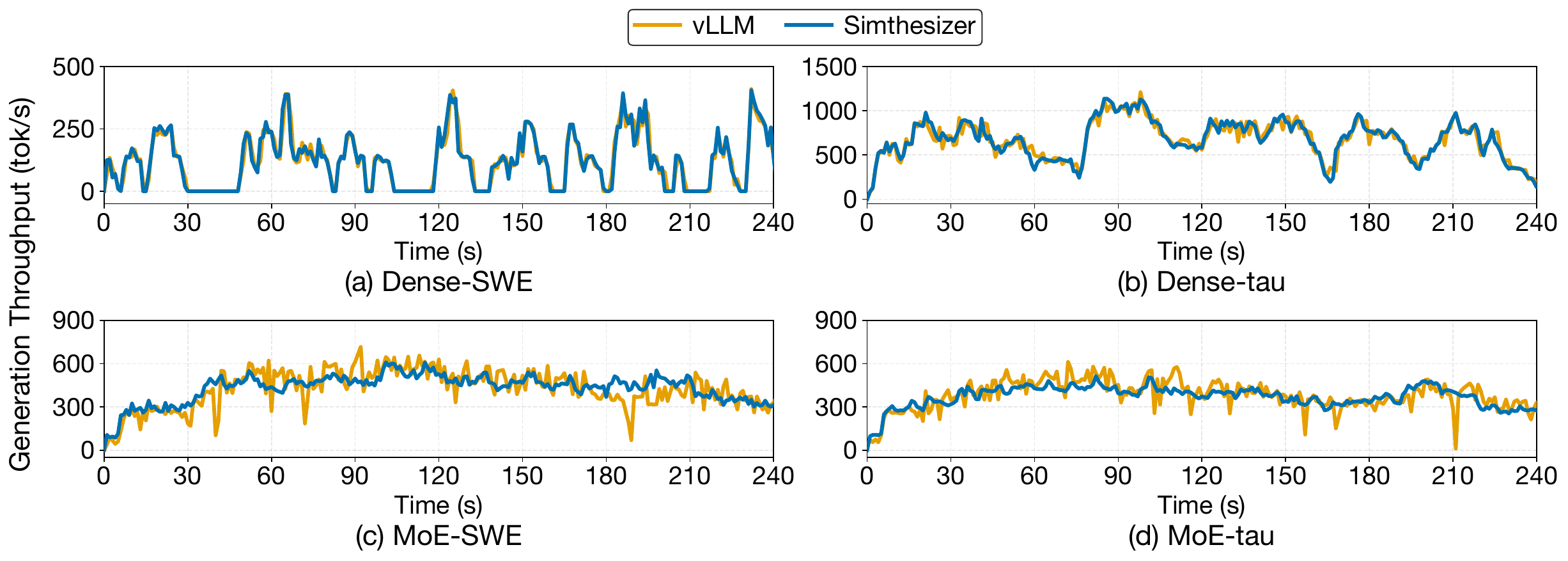}
    \vspace{-1ex}
    \caption{Comparison of throughput between unmodified \sysname{} (without any agent-synthesized extension) and a GPU-based serving system running vLLM under agentic workloads: (a) Llama3.1 dense model on mini-SWE-bench, (b) Llama3.1 dense model on tau-bench, (c) Qwen3 MoE model on mini-SWE-bench, and (d) Qwen3 MoE model on tau-bench.}
    \vspace{-1ex}
    \label{fig:agent-throughput-validation}
\end{figure*}

\subsection{Resolving Simulator Semantics}
The initial user request states target algorithmic details including
the number of draft tokens and the acceptance model, but leaves some
modeling decisions unspecified.
Based on the reference and the supplied statistics,
\texttt{task-design} resolves these decisions into explicit simulator semantics.
For example, \synthesizer{} derives four semantic variables from the
algorithm reference: the verification width \(W\), the remaining
output length \(R\), the accepted draft length \(A\), and the
committed request progress \(M\).
Figure~\ref{fig:case-study-specification} presents the example
specification document generated by \texttt{task-design} from an
underspecified user request, capturing the resulting modeling
decisions and state semantics.

\subsection{From Mapping to Implementation}
\texttt{sim-mapping} maps the simulator-facing specification onto the
 \substrate{} components and interfaces, recording the
resulting mappings in an implementation map for subsequent simulator
development.
Rather than adding mechanism-specific machinery, \synthesizer{}
localizes extension logic within existing \substrate{} components and
minimizes the required changes.
For example, in speculative decoding, the map incorporates simulator semantics, $W$ and $M$, into the existing request-processing and scheduling mechanisms, while reusing the common DAG scheduling and execution path.
Table~\ref{tab:spec-mapping} summarizes this resulting map.

During the \texttt{sim-dev} stage, \synthesizer{} translates the
mapped design into localized simulator code changes guided by the
implementation map.
The resulting extension introduces no new events, node types, graph
APIs, model-runner paths, compute interfaces, cache APIs, or
execution loops, leaving most \substrate{} components unchanged.
Overall, \sysname{}'s harnessed lowering process allows
\synthesizer{} to focus on the algorithmic details and state
semantics specific to speculative decoding, while \substrate{}
supports them through its reusable components, cost models, and interfaces.

\begin{figure*}[t]
    \centering
    \includegraphics[width=0.9\linewidth]{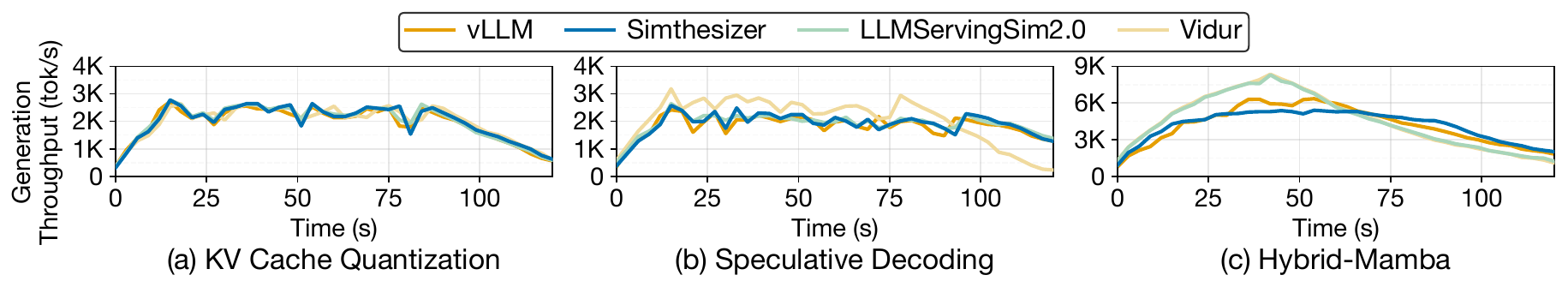}
    \vspace{-1ex}
    \caption{Throughput comparison of \sysname{}, LLMServingSim2.0, and Vidur each independently extended by the Synthesizer agent under a shared harness and coding agent across three feature-extension tasks: (a) KV cache quantization, (b) speculative decoding, and (c) hybrid Mamba model support. Each panel thus compares three distinct simulators extended with the same feature; the same convention applies to Fig.~\ref{fig:error-rate-comparison} and Fig.~\ref{fig:simulation-time-comparison}.}
    \vspace{-1ex}
    \label{fig:baseline-throughput-comparison}
\end{figure*}

\subsection{Validation-Driven Revision}
The synthesis run produces an executable candidate simulator.
In this representative run, trace-guided validation found that the
candidate overestimated target system throughput by \(13.4\%\) on the
validation trace.
Inspecting its specification and computation trace revealed that the
synthesized implementation applied \(W\) only to the final layer and
modeled the remaining target model layers over \(K\) positions,
causing the candidate to underestimate the required compute and communication.
The validation evidence isolated the problem to the
verification-width semantics rather than the component mapping or DAG
execution path.

The validation report fed this finding back to \synthesizer{}, which
initiated another iteration to correct the semantic error.
With the corrected semantics, the subsequent candidate showed a
\(6.7\%\) throughput divergence from the real system and
underestimated aggregate throughput only by \(4.7\%\) on the same
validation trace.
This case study shows how trace-guided validation converts a
simulation discrepancy into a semantic correction while
preserving the localized integration design.

\subsection{Joint Effect}
\substrate{} provides reusable components for scheduling, compute,
cache management, and DAG execution, while \synthesizer{} maps
underspecified behavior onto these components.
In this speculative decoding case, this combination localizes
mechanism-specific logic to the acceptance policy and scheduler,
preserves shared execution and resource-modeling paths, and links
observed performance discrepancies to concrete modeling decisions.
Together, the two components of \sysname{} make simulator extension
compositional and reviewable, exposing each extension's semantics,
integration choices, and validation evidence to human review.

\section{Evaluation}
\label{sec:evaluation}

Our evaluation answers five questions:
\textbf{(Q1)} Can \sysname{} accurately model complex workloads such as agent serving?
\textbf{(Q2)} Can \sysname{} faithfully implement features not natively supported by \substrate{}, and how much does \substrate{} contribute to simulation accuracy?
\textbf{(Q3)} How much does the \synthesizer{} harness contribute to accuracy?
\textbf{(Q4)} How fast does \sysname{} run LLM serving simulations?
\textbf{(Q5)} Does the workflow remain effective across coding agents?

\subsection{Methodology}
\niparagraph{Baselines.}
We compare \sysname{} against two state-of-the-art LLM serving simulators: LLMServingSim2.0~\cite{llmservingsim2} and Vidur~\cite{vidur}.
To evaluate agent-driven simulator extension, we define three feature-extension tasks unsupported by \substrate{} and both baselines: (1) FP8 KV cache quantization, (2) EAGLE3-based speculative decoding~\cite{eagle3}, and (3) hybrid Mamba model support.
We use the Qwen3 32B model~\cite{qwen3} for the first two tasks and the Nemotron-3-Nano 30B-A3B model~\cite{nemotron3} for the hybrid Mamba task.
Within \sysname{}, \synthesizer{} adds each requested feature directly to \substrate{}; for each baseline, the same coding agent and harness add the same mechanism to the simulator's native implementation.
We repeat each extension task over multiple independent trials without human intervention and report metrics averaged across trials to account for coding-agent variability.

\niparagraph{\synthesizer{} setup.}
Unless otherwise specified, we use GPT-5.4 (xhigh), accessed through OpenAI Codex (v0.118.0), as \synthesizer{}.
Importantly, the \synthesizer{} harness does not encode any \sysname{}-specific information.
For fair comparison, we provide all three simulators with identical prompts, references, specifications, and profiled model data.

\niparagraph{Workloads and datasets.}
We sample 50--100 requests from SWE-bench~\cite{swe-bench} and tau-bench~\cite{taubench}, which are agentic workloads, and we sample 400 requests from ShareGPT~\cite{sharegpt} as a non-agentic workload.
For agentic workloads, we collect output and tool-call trajectories from a real system matching the simulated environment and replay them in the simulator.

\niparagraph{System specifications.}
We conduct all experiments on a machine equipped with two NVIDIA RTX Pro 6000 Blackwell GPUs and an Intel Xeon Gold 6326 CPU.
We use vLLM (v0.19.1) as the LLM serving framework when evaluating the real GPU-based serving system~\cite{vllm}.

\subsection{Evaluation Results}
\begin{table}[t]
    \caption{Error rates of \sysname{} performance metrics relative to the real serving system across two workloads and two models.}
    \centering
    \includegraphics[width=1.0\linewidth]{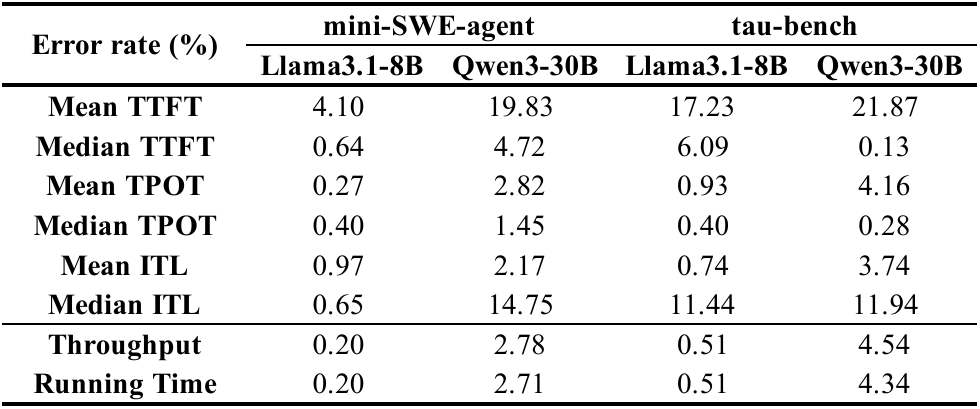}
    \label{tab:agent-error} 
    \vspace{-3em}
\end{table}

\niparagraph{(Q1) Complex-workload fidelity.}
We evaluate whether \sysname{} captures dynamic and complex workload behavior by comparing it against a real GPU-based serving system running vLLM on mini-SWE-bench and tau-bench.
Both benchmarks define their agentic workflows in a pre-defined form, which \sysname{} models directly as multiple model invocations interleaved with tool calls.
For each tool call, we execute it on the host in advance and record its latency, which the simulation replays at the corresponding stage.
Figure~\ref{fig:agent-throughput-validation} illustrates the generation throughput over 240 seconds for the Llama3.1 8B dense model~\cite{llama3} and the Qwen3 30B-A3B MoE model~\cite{qwen3}.

For the dense model (Figure~\ref{fig:agent-throughput-validation}(a) and (b)), the simulation closely matches the real system, as dense execution is largely deterministic.
For the MoE model (Figure~\ref{fig:agent-throughput-validation}(c) and (d)), the simulated throughput deviates more visibly.
This discrepancy arises from the statistical modeling of MoE routing; expert selection and synchronization under data and expert parallelism are estimated from profiled statistics rather than actual request-specific routing decisions.
Nevertheless, \sysname{} captures the interaction between LLM generation and tool calls and closely follows the real system's performance trends.

Table~\ref{tab:agent-error} reports the error rates of \sysname{} for TTFT, TPOT, and ITL.
TTFT and ITL exhibit higher errors than TPOT because real systems involve tail-latency factors, such as network traffic and kernel launch delays, that are difficult to simulate precisely.
TPOT, which primarily reflects generation-phase latency, shows lower error, indicating that \sysname{} accurately models batching and scheduling, where these system-level effects are less dominant.
Despite these hard-to-simulate system-level effects, \sysname{} closely reproduces the real system's overall behavior across all metrics, workloads, and models.

\begin{figure}[t]
    \centering
    \includegraphics[width=0.9\linewidth]{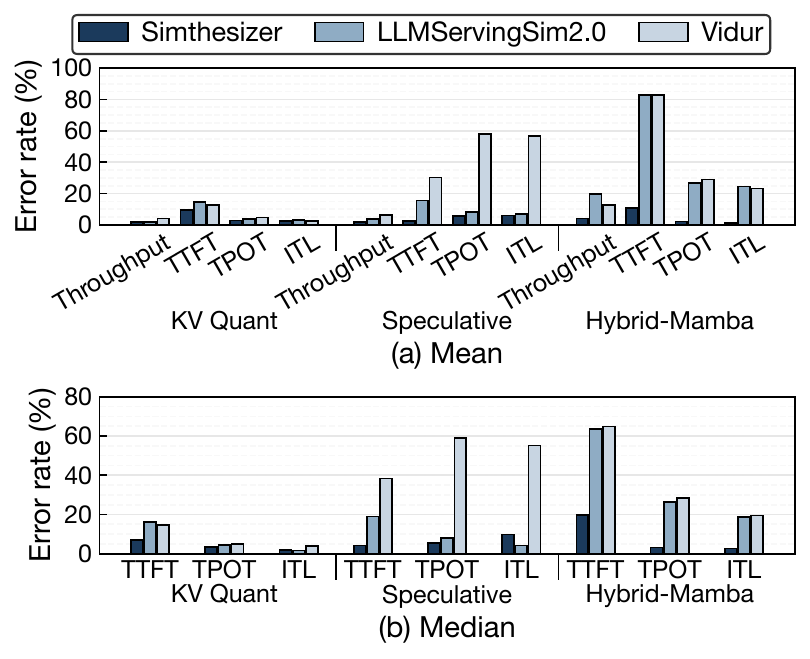}
    \vspace{-1ex}
    \caption{Error rate comparison of (a) mean and (b) median performance metrics for \sysname{}, LLMServingSim2.0, and Vidur across three feature-extension tasks under a shared harness and coding agent.}
    \vspace{-1ex}
    \label{fig:error-rate-comparison}
\end{figure}

\niparagraph{(Q2) Extension fidelity.}
We next evaluate the three feature-extension tasks that \synthesizer{} adds to \substrate{} and to each baseline simulator.
Figure~\ref{fig:baseline-throughput-comparison} presents the generation throughput, and Figure~\ref{fig:error-rate-comparison} shows the error rates of key performance metrics across the three simulators.

\begin{table}[t]
    \caption{Error rates of \sysname{} with and without the \synthesizer{} harness, relative to the real vLLM-based serving system.}
    \centering
    \includegraphics[width=1.0\linewidth]{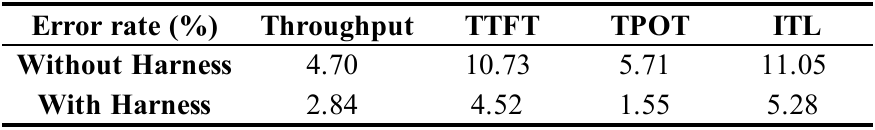}
    \label{tab:agent-harness-error} 
    \vspace{-3em}
\end{table}

For KV cache quantization, all simulators achieve low error rates because the task primarily requires tracking memory usage and modeling quantization overhead.
In contrast, speculative decoding and hybrid Mamba support require coordinated changes across execution scheduling, model structure, and performance modeling.
For these two tasks, \sysname{} consistently achieves the highest accuracy, whereas the baselines exhibit substantially larger errors.
This difference arises because \substrate{}'s modular interfaces localize each extension, while the fixed pipelines of existing simulators require changes across tightly coupled components to represent interactions among Mamba, MoE, and transformer layers.

Because the coding agent, harness, specifications, and profiling data are identical across all three systems, the accuracy difference isolates the contribution of the underlying simulator.
Overall, \sysname{} achieves an average throughput error of 2.51\%, whereas the baseline-built extensions exhibit 6.03\%, showing that \sysname{} faithfully adds functionality absent from \substrate{} and that \substrate{} provides a more effective foundation for agent-driven extension.

\niparagraph{(Q3) Impact of the \synthesizer{} harness.}
Table~\ref{tab:agent-harness-error} compares \sysname{} after \synthesizer{} adds speculative decoding with and without its harness, using the ShareGPT workload and the Qwen3 32B model.
Without the harness, simulation error rises by 1.65$\times$--3.69$\times$ across throughput and latency metrics.
This occurs because the coding agent frequently omits critical system details and proceeds to implementation despite insufficient specifications.
In contrast, the harness supplies a knowledge base, blocks implementation until required specifications are resolved, and enforces self-validation through its verification loop.
Because this comparison holds \substrate{}, the coding agent, and all inputs fixed, the error reduction quantifies the harness's contribution to accuracy.

\begin{figure}[t]
    \centering
    \includegraphics[width=0.9\linewidth]{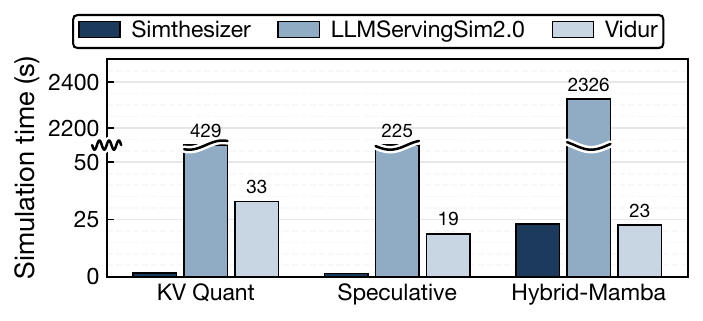}
    \vspace{-1ex}
    \caption{Comparison of simulation time for \sysname{}, LLMServingSim2.0, and Vidur across three feature-extension tasks.}
    \vspace{-2ex}
    \label{fig:simulation-time-comparison}
\end{figure}

\begin{figure}[t]
    \centering
    \includegraphics[width=0.9\linewidth]{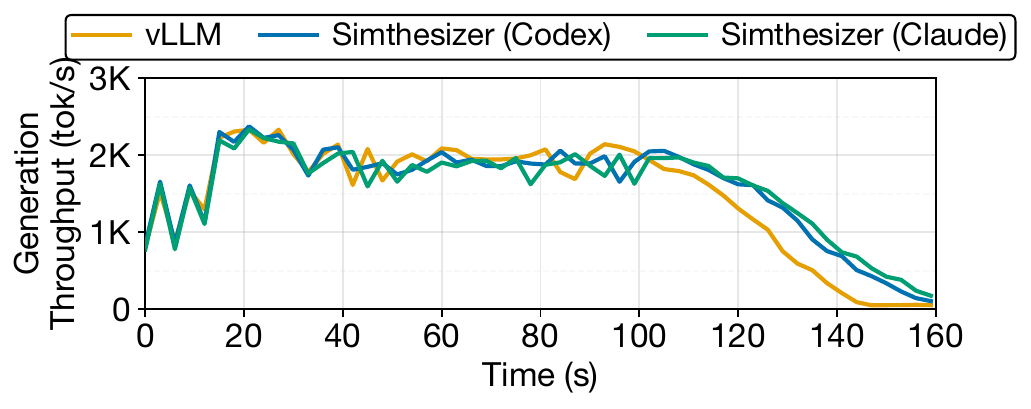}
    \vspace{-2ex}
    \caption{Generation throughput comparison between vLLM and the simulators produced by Codex and Claude Code for speculative decoding.}
    \vspace{-2ex}
    \label{fig:codex-vs-claude}
\end{figure}

\niparagraph{(Q4) Simulation speed.}
Figure~\ref{fig:simulation-time-comparison} compares the execution time of \sysname{}, LLMServingSim2.0, and Vidur across the three extension tasks under the ShareGPT workload.
\sysname{} runs on average 164.9$\times$ (up to 284.96$\times$) faster than LLMServingSim2.0 and 6.65$\times$ (up to 23.19$\times$) faster than Vidur.
This gap arises from how each extension is implemented.
Existing simulators expose no modular boundaries for new mechanisms, so extension logic is built directly into their simulation pipelines, adding overhead to every simulated iteration.
In contrast, on \sysname{}, the same extensions remain confined to localized component changes behind \substrate{}'s interfaces, leaving the execution engine untouched.
For hybrid Mamba, \sysname{} takes longer than on the other tasks because \substrate{} normally simulates one transformer layer and reuses its latency across repeated layers, whereas the heterogeneous layers of hybrid Mamba leads \substrate{} to simulate every layer individually.
Explicit request of this reuse would further reduce the simulation time, yet even without it, \sysname{} models hybrid Mamba more accurately than both baselines.

\niparagraph{(Q5) Coding-agent generality.}
Figure~\ref{fig:codex-vs-claude} compares the generation throughput of vLLM with that of the simulators produced by Codex and Claude Code (Opus 4.7 high, v2.1.204) for speculative decoding.
Both simulators closely follow the real-system trajectory, achieving average error rates of 4.1\% and 4.9\%, respectively, across throughput, TTFT, TPOT, and ITL.
This generality follows from the agent- and LLM-agnostic harness design, which relies on no agent-specific tools or prompting strategies and applies the same specification-to-implementation and validation workflow to any coding agent.
These results show that \synthesizer{} generalizes beyond Codex, guiding different coding agents to faithful extensions.

\section{Related Work}
\niparagraph{LLM and system simulators.}
ADOR~\cite{ador}, ONNXim~\cite{onnxim}, LLMCompass~\cite{llmcompass}, and PyTorchSim~\cite{pytorchsim} simulate LLM workloads at the hardware level to explore architectural design spaces.
At the system level, Vidur~\cite{vidur}, APEX~\cite{apex}, and LLMServingSim2.0~\cite{llmservingsim2} model LLM serving across nodes and devices, while vTrain~\cite{vTrain}, TrioSim~\cite{triosim}, and Multiverse~\cite{multiverse} target distributed training.
Across all these levels, however, human developers implement system behavior in advance and fix it into a simulation pipeline, so every emerging mechanism triggers another round of manual restructuring and revalidation.
In contrast, \sysname{} composes the complete serving workflow from uniform elements of a unified dynamic DAG, and integrates new mechanisms by adding and connecting nodes through agent-driven lowering.

\niparagraph{AI agents for systems research.}
KNighter~\cite{knighter} synthesizes static-analysis checkers, AIOpsLab~\cite{aiopslab} evaluates agents for autonomous cloud operation, LOOPRAG~\cite{looprag} and LLM-Vectorizer~\cite{llm-vectorizer} vectorize loops, and CUDAForge~\cite{zhang2025cudaforge} and KernelEvolve~\cite{kernel_evolve} generate and optimize GPU kernels.
\sysname further demonstrates that coding agents can be applied to develop an LLM serving simulator, showing the feasibility of agent-based development and optimization through artifacts and workflows co-designed for agents.

\section{Conclusion}
\label{sec:conclusion}

This paper presents \sysname{}, a framework for developing LLM serving simulators that keep pace with fast-evolving, non-monolithic serving.
\sysname{} introduces a composable simulator infrastructure that uniformly expresses the complete serving workflow, including its control decisions, realized as a unified dynamic DAG (\substrate{}), and lowers high-level specifications into executable extensions through a harnessed coding agent (\synthesizer{}).
Extensions synthesized on \sysname{} achieve an average throughput error of 2.51\%, compared with 6.03\% for the same extensions built on existing simulators, while simulating up to 23.19$\times$--284.96$\times$ faster.
These results suggest that pairing composable simulator infrastructure with agent-driven lowering offers a scalable path toward building simulators in other rapidly evolving systems domains.

\bibliographystyle{IEEEtranS}
\bibliography{references}

\begin{thebibliography}{10}
\providecommand{\url}[1]{#1}
\csname url@samestyle\endcsname
\providecommand{\newblock}{\relax}
\providecommand{\bibinfo}[2]{#2}
\providecommand{\BIBentrySTDinterwordspacing}{\spaceskip=0pt\relax}
\providecommand{\BIBentryALTinterwordstretchfactor}{4}
\providecommand{\BIBentryALTinterwordspacing}{\spaceskip=\fontdimen2\font plus
\BIBentryALTinterwordstretchfactor\fontdimen3\font minus
  \fontdimen4\font\relax}
\providecommand{\BIBforeignlanguage}[2]{{%
\expandafter\ifx\csname l@#1\endcsname\relax
\typeout{** WARNING: IEEEtranS.bst: No hyphenation pattern has been}%
\typeout{** loaded for the language `#1'. Using the pattern for}%
\typeout{** the default language instead.}%
\else
\language=\csname l@#1\endcsname
\fi
#2}}
\providecommand{\BIBdecl}{\relax}
\BIBdecl

\bibitem{vidur}
A.~Agrawal, N.~Kedia, J.~Mohan, A.~Panwar, N.~Kwatra, B.~S. Gulavani,
  R.~Ramjee, and A.~Tumanov, ``{VIDUR: A LARGE-SCALE SIMULATION FRAMEWORK FOR
  LLM INFERENCE},'' in \emph{MLSys}, 2024.

\bibitem{sarathi-serve}
A.~Agrawal, N.~Kedia, A.~Panwar, J.~Mohan, N.~Kwatra, B.~S. Gulavani,
  A.~Tumanov, and R.~Ramjee, ``{Taming Throughput-Latency Tradeoff in LLM
  Inference with Sarathi-Serve},'' \emph{OSDI}, 2024.

\bibitem{vTrain}
J.~Bang, Y.~Choi, M.~Kim, Y.~Kim, and M.~Rhu, ``{vTrain: A Simulation Framework
  for Evaluating Cost-Effective and Compute-Optimal Large Language Model
  Training},'' in \emph{MICRO}, 2024.

\bibitem{aiopslab}
Y.~Chen, M.~Shetty, G.~Somashekar, M.~Ma, Y.~Simmhan, J.~Mace, C.~Bansal,
  R.~Wang, and S.~Rajmohan, ``{AIOpsLab: A Holistic Framework to Evaluate AI
  Agents for Enabling Autonomous Clouds},'' in \emph{MLSys}, 2025.

\bibitem{pascal}
E.~Cho, J.~Bang, R.~Hwang, and M.~Rhu, ``{ PASCAL: A Phase-Aware Scheduling
  Algorithm for Serving Reasoning-based Large Language Models},'' in
  \emph{HPCA}, 2026.

\bibitem{llmservingsim2}
J.~Cho, H.~Choi, G.~Heo, and J.~Park, ``{LLMServingSim 2.0: A Unified Simulator
  for Heterogeneous and Disaggregated LLM Serving Infrastructure},'' in
  \emph{ISPASS}, 2026.

\bibitem{llmservingsim}
J.~Cho, M.~Kim, H.~Choi, G.~Heo, and J.~Park, ``{LLMServingSim: A HW/SW
  Co-Simulation Infrastructure for LLM Inference Serving at Scale},'' in
  \emph{IISWC}, 2024.

\bibitem{cached_attention}
B.~Gao, Z.~He, P.~Sharma, Q.~Kang, D.~Jevdjic, J.~Deng, X.~Yang, Z.~Yu, and
  P.~Zuo, ``{{Cost-Efficient} Large Language Model Serving for Multi-turn
  Conversations with {CachedAttention}},'' in \emph{ATC}, 2024.

\bibitem{pie}
I.~Gim, Z.~Ma, S.-s. Lee, and L.~Zhong, ``{Pie: A Programmable Serving System
  for Emerging LLM Applications},'' in \emph{SOSP}, 2025.

\bibitem{multiverse}
F.~Gui, K.~Gao, L.~Chen, D.~Li, V.~Liu, R.~Zhang, H.~Yang, and D.~Xiong,
  ``{Accelerating design space exploration for LLM training systems with
  multi-experiment parallel simulation},'' in \emph{NSDI}, 2025.

\bibitem{onnxim}
H.~Ham, W.~Yang, Y.~Shin, O.~Woo, G.~Heo, S.~Lee, J.~Park, and G.~Kim,
  ``{ONNXim: A Fast, Cycle-Level Multi-Core NPU Simulator},'' \emph{IEEE
  Computer Architecture Letters}, vol.~23, no.~2, pp. 219--222, 2024.

\bibitem{ns3}
T.~R. Henderson, M.~Lacage, G.~F. Riley, C.~Dowell, and J.~Kopena, ``{Network
  simulations with the ns-3 simulator},'' \emph{SIGCOMM demonstration},
  vol.~14, no.~14, p. 527, 2008.

\bibitem{metagpt}
S.~Hong, M.~Zhuge, J.~Chen, X.~Zheng, Y.~Cheng, J.~Wang, C.~Zhang, Z.~Wang,
  S.~K.~S. Yau, Z.~Lin, L.~Zhou, C.~Ran, L.~Xiao, C.~Wu, and J.~Schmidhuber,
  ``{Meta{GPT}: Meta Programming for A Multi-Agent Collaborative Framework},''
  in \emph{ICLR}, 2024.

\bibitem{flashgen}
J.~Jeong and J.~Ahn, ``{Accelerating LLM Serving for Multi-turn Dialogues with
  Efficient Resource Management},'' in \emph{ASPLOS}, 2025.

\bibitem{swe-bench}
C.~E. Jimenez, J.~Yang, A.~Wettig, S.~Yao, K.~Pei, O.~Press, and K.~Narasimhan,
  ``Swe-bench: Can language models resolve real-world github issues?'' in
  \emph{International Conference on Learning Representations}, vol. 2024, 2024,
  pp. 54\,107--54\,157.

\bibitem{asplos25_pod_attention}
\BIBentryALTinterwordspacing
A.~K. Kamath, R.~Prabhu, J.~Mohan, S.~Peter, R.~Ramjee, and A.~Panwar,
  ``{POD-Attention: Unlocking Full Prefill-Decode Overlap for Faster LLM
  Inference},'' in \emph{ASPLOS}, 2025. [Online]. Available:
  \url{https://asplos-conference.org/asplos2025/program.html}
\BIBentrySTDinterwordspacing

\bibitem{accel_sim}
M.~Khairy, Z.~Shen, T.~M. Aamodt, and T.~G. Rogers, ``{Accel-Sim: An Extensible
  Simulation Framework for Validated GPU Modeling},'' in \emph{ISCA}, 2020.

\bibitem{ador}
J.~Kim, H.~Lee, G.~Ko, G.~Choi, S.~Ham, S.~Hong, and J.-Y. Kim, ``{ADOR: A
  Design Exploration Framework for LLM Serving with Enhanced Latency and
  Throughput},'' in \emph{ISPASS}, 2025.

\bibitem{pimba}
W.~Kim, Y.~Lee, Y.~Kim, J.~Hwang, S.~Oh, J.~Jung, A.~Huseynov, W.~G. Park,
  C.~H. Park, D.~Mahajan, and J.~Park, ``{Pimba: A Processing-in-Memory
  Acceleration for Post-Transformer Large Language Model Serving},'' in
  \emph{MICRO}, 2025.

\bibitem{vllm}
W.~Kwon, Z.~Li, S.~Zhuang, Y.~Sheng, L.~Zheng, C.~H. Yu, J.~Gonzalez, H.~Zhang,
  and I.~Stoica, ``{Efficient Memory Management for Large Language Model
  Serving with PagedAttention},'' in \emph{SOSP}, 2023.

\bibitem{spec_dec}
Y.~Leviathan, M.~Kalman, and Y.~Matias, ``{Fast Inference from Transformers via
  Speculative Decoding},'' in \emph{ICML}, 2023.

\bibitem{triosim}
Y.~Li, Y.~Bao, G.~Wang, X.~Mei, P.~Vaid, A.~Ghosh, A.~Jog, D.~Bunandar,
  A.~Joshi, and Y.~Sun, ``{TrioSim: A Lightweight Simulator for Large-Scale DNN
  Workloads on Multi-GPU Systems},'' in \emph{ISCA}, 2025.

\bibitem{eagle2}
Y.~Li, F.~Wei, C.~Zhang, and H.~Zhang, ``{{EAGLE}-2: Faster Inference of
  Language Models with Dynamic Draft Trees},'' in \emph{EMNLP}, 2024.

\bibitem{eagle1}
Y.~Li, F.~Wei, C.~Zhang, and H.~Zhang, ``{{EAGLE}: Speculative Sampling
  Requires Rethinking Feature Uncertainty},'' in \emph{ICML}, 2024.

\bibitem{eagle3}
Y.~Li, F.~Wei, C.~Zhang, and H.~Zhang, ``{{EAGLE}-3: Scaling up Inference
  Acceleration of Large Language Models via Training-Time Test},'' in
  \emph{NeurIPS}, 2025.

\bibitem{kernel_evolve}
\BIBentryALTinterwordspacing
G.~Liao, H.~Qin, Y.~Wang, A.~Golden, M.~Kuchnik, Y.~Yetim, J.~J. Ang, C.~Fu,
  Y.~He, S.~Hsia, Z.~Jiang, D.~Li, U.~Pashkevich, V.~Puvvada, F.~Shi,
  M.~Steiner, R.~Xiao, N.~Yan, X.~Yu, Z.~Fang, R.~Levenstein, K.~Ho, H.~Zhu,
  A.~Hammond, R.~Li, A.~Mathews, K.~Gondkar, A.~Zainul-Abedin, K.~Singh, H.~Yu,
  W.~Chi, B.~Huang, S.~Zhang, N.~Weller, Z.~Marine, W.~Cook, C.-J. Wu, and
  G.~Liu, ``{KernelEvolve: Scaling Agentic Kernel Coding for Heterogeneous AI
  Accelerators at Meta},'' 2026. [Online]. Available:
  \url{https://arxiv.org/abs/2512.23236}
\BIBentrySTDinterwordspacing

\bibitem{parrot}
C.~Lin, Z.~Han, C.~Zhang, Y.~Yang, F.~Yang, C.~Chen, and L.~Qiu, ``{Parrot:
  Efficient Serving of {LLM-based} Applications with Semantic Variable},'' in
  \emph{OSDI}, 2024.

\bibitem{apex}
\BIBentryALTinterwordspacing
Y.-C. Lin, W.~Kwon, R.~Pineda, and F.~N. Paravecino, ``{APEX: An Extensible and
  Dynamism-Aware Simulator for Automated Parallel Execution in LLM Serving},''
  2025. [Online]. Available: \url{https://arxiv.org/abs/2411.17651}
\BIBentrySTDinterwordspacing

\bibitem{llama3}
{Meta AI}, ``{Llama 3.1 8B Instruct},''
  \url{https://huggingface.co/meta-llama/Llama-3.1-8B-Instruct}, Jul. 2024,
  accessed: 2026-04-09.

\bibitem{specinfer}
X.~Miao, G.~Oliaro, Z.~Zhang, X.~Cheng, Z.~Wang, Z.~Zhang, R.~Y.~Y. Wong,
  A.~Zhu, L.~Yang, X.~Shi, C.~Shi, Z.~Chen, D.~Arfeen, R.~Abhyankar, and
  Z.~Jia, ``{SpecInfer: Accelerating Large Language Model Serving with
  Tree-based Speculative Inference and Verification},'' in \emph{ASPLOS}, 2024.

\bibitem{nemotron3}
{NVIDIA Corporation}, ``{NVIDIA Nemotron-3 Nano 30B A3B BF16},''
  \url{https://huggingface.co/nvidia/NVIDIA-Nemotron-3-Nano-30B-A3B-BF16}, Dec.
  2025, accessed: 2026-04-09.

\bibitem{suffix_decoding}
G.~Oliaro, Z.~Jia, D.~F. Campos, and A.~Qiao, ``{SuffixDecoding: Extreme
  Speculative Decoding for Emerging {AI} Applications},'' in \emph{NeurIPS},
  2025.

\bibitem{codex}
{OpenAI}, ``{Codex},'' \url{https://openai.com/codex/}, 2025.

\bibitem{splitwise}
P.~Patel, E.~Choukse, C.~Zhang, A.~Shah, I.~n. Goiri, S.~Maleki, and
  R.~Bianchini, ``{Splitwise: Efficient Generative LLM Inference Using Phase
  Splitting},'' in \emph{ISCA}, 2024.

\bibitem{vattention}
R.~Prabhu, A.~Nayak, J.~Mohan, R.~Ramjee, and A.~Panwar, ``{vAttention: Dynamic
  Memory Management for Serving LLMs without PagedAttention},'' in
  \emph{ASPLOS}, 2025.

\bibitem{astra_sim}
S.~Rashidi, S.~Sridharan, S.~Srinivasan, and T.~Krishna, ``{ASTRA-SIM: Enabling
  SW/HW Co-Design Exploration for Distributed DL Training Platforms},'' in
  \emph{ISPASS}, 2020.

\bibitem{toolformer}
T.~Schick, J.~Dwivedi-Yu, R.~Dessi, R.~Raileanu, M.~Lomeli, E.~Hambro,
  L.~Zettlemoyer, N.~Cancedda, and T.~Scialom, ``{Toolformer: Language Models
  Can Teach Themselves to Use Tools},'' in \emph{NeurIPS}, 2023.

\bibitem{sharegpt}
{ShareGPT Team}, ``{ShareGPT},'' \url{https://sharegpt.com}, 2023, accessed:
  2026-04-09.

\bibitem{hugging_gpt}
Y.~Shen, K.~Song, X.~Tan, D.~Li, W.~Lu, and Y.~Zhuang, ``{HuggingGPT: Solving
  AI Tasks with ChatGPT and its Friends in Hugging Face},'' in \emph{NeurIPS},
  2023.

\bibitem{reflexion}
N.~Shinn, F.~Cassano, A.~Gopinath, K.~Narasimhan, and S.~Yao, ``{Reflexion:
  language agents with verbal reinforcement learning},'' in \emph{NeurIPS},
  2023.

\bibitem{llumnix}
B.~Sun, Z.~Huang, H.~Zhao, W.~Xiao, X.~Zhang, Y.~Li, and W.~Lin, ``{Llumnix:
  Dynamic Scheduling for Large Language Model Serving},'' in \emph{OSDI}, 2024.

\bibitem{mgpusim}
Y.~Sun, T.~Baruah, S.~A. Mojumder, S.~Dong, X.~Gong, S.~Treadway, Y.~Bao,
  S.~Hance, C.~McCardwell, V.~Zhao, H.~Barclay, A.~K. Ziabari, Z.~Chen,
  R.~Ubal, J.~L. Abell\'{a}n, J.~Kim, A.~Joshi, and D.~Kaeli, ``{MGPUSim:
  enabling multi-GPU performance modeling and optimization},'' in \emph{ISCA},
  2019.

\bibitem{llm-vectorizer}
J.~Taneja, A.~Laird, C.~Yan, M.~Musuvathi, and S.~K. Lahiri, ``{LLM-Vectorizer:
  LLM-Based Verified Loop Vectorizer},'' in \emph{CGO}, 2025.

\bibitem{kvcache_cache}
J.~Wang, J.~Han, X.~Wei, S.~Shen, D.~Zhang, C.~Fang, R.~Chen, W.~Yu, and
  H.~Chen, ``{$\{$KVCache$\}$ Cache in the Wild: Characterizing and Optimizing
  $\{$KVCache$\}$ Cache at a Large Cloud Provider},'' in \emph{ATC}, 2025.

\bibitem{loop_serve}
B.~Wu, S.~Liu, Y.~Zhong, P.~Sun, X.~Liu, and X.~Jin, ``{LoongServe: Efficiently
  Serving Long-Context Large Language Models with Elastic Sequence
  Parallelism},'' in \emph{SOSP}, 2024.

\bibitem{autogen}
Q.~Wu, G.~Bansal, J.~Zhang, Y.~Wu, B.~Li, E.~Zhu, L.~Jiang, X.~Zhang, S.~Zhang,
  J.~Liu, A.~H. Awadallah, R.~W. White, D.~Burger, and C.~Wang, ``{AutoGen:
  Enabling Next-Gen {LLM} Applications via Multi-Agent Conversations},'' in
  \emph{COLM}, 2024.

\bibitem{qwen3}
\BIBentryALTinterwordspacing
A.~Yang, A.~Li, B.~Yang, B.~Zhang, B.~Hui, B.~Zheng, B.~Yu, C.~Gao, C.~Huang,
  C.~Lv, C.~Zheng, D.~Liu, F.~Zhou, F.~Huang, F.~Hu, H.~Ge, H.~Wei, H.~Lin,
  J.~Tang, J.~Yang, J.~Tu, J.~Zhang, J.~Yang, J.~Yang, J.~Zhou, J.~Zhou,
  J.~Lin, K.~Dang, K.~Bao, K.~Yang, L.~Yu, L.~Deng, M.~Li, M.~Xue, M.~Li,
  P.~Zhang, P.~Wang, Q.~Zhu, R.~Men, R.~Gao, S.~Liu, S.~Luo, T.~Li, T.~Tang,
  W.~Yin, X.~Ren, X.~Wang, X.~Zhang, X.~Ren, Y.~Fan, Y.~Su, Y.~Zhang, Y.~Zhang,
  Y.~Wan, Y.~Liu, Z.~Wang, Z.~Cui, Z.~Zhang, Z.~Zhou, and Z.~Qiu, ``{Qwen3
  Technical Report},'' 2025. [Online]. Available:
  \url{https://arxiv.org/abs/2505.09388}
\BIBentrySTDinterwordspacing

\bibitem{knighter}
C.~Yang, Z.~Zhao, Z.~Xie, H.~Li, and L.~Zhang, ``{KNighter: Transforming Static
  Analysis with LLM-Synthesized Checkers},'' in \emph{SOSP}, 2025.

\bibitem{pytorchsim}
W.~Yang, Y.~Shin, O.~Woo, G.~Park, H.~Ham, J.~Kang, J.~Park, and G.~Kim,
  ``{PyTorchSim: A Comprehensive, Fast, and Accurate NPU Simulation
  Framework},'' in \emph{MICRO}, 2025.

\bibitem{taubench}
S.~Yao, N.~Shinn, P.~Razavi, and K.~R. Narasimhan,
  ``{$\tau$-bench: A Benchmark for
  \underline{T}ool-\underline{A}gent-\underline{U}ser Interaction in Real-World
  Domains},'' in \emph{ICLR}, 2025.

\bibitem{react}
S.~Yao, J.~Zhao, D.~Yu, N.~Du, I.~Shafran, K.~R. Narasimhan, and Y.~Cao,
  ``{ReAct: Synergizing Reasoning and Acting in Language Models},'' in
  \emph{ICLR}, 2023.

\bibitem{deltazip}
X.~Yao, Q.~Hu, and A.~Klimovic, ``{DeltaZip: Efficient Serving of Multiple
  Full-Model-Tuned LLMs},'' in \emph{EuroSys}, 2025.

\bibitem{orca}
G.-I. Yu, J.~S. Jeong, G.-W. Kim, S.~Kim, and B.-G. Chun, ``{Orca: A
  Distributed Serving System for {Transformer-Based} Generative Models},'' in
  \emph{OSDI}, 2022.

\bibitem{pensieve}
L.~Yu, J.~Lin, and J.~Li, ``{Stateful Large Language Model Serving with
  Pensieve},'' in \emph{EuroSys}, 2025.

\bibitem{llmcompass}
H.~Zhang, A.~Ning, R.~B. Prabhakar, and D.~Wentzlaff, ``{LLMCompass: Enabling
  Efficient Hardware Design for Large Language Model Inference},'' in
  \emph{ISCA}, 2025.

\bibitem{zhang2025cudaforge}
\BIBentryALTinterwordspacing
Z.~Zhang, R.~Wang, S.~Li, Y.~Luo, M.~Hong, and C.~Ding, ``{{CudaForge}: An
  Agent Framework with Hardware Feedback for {CUDA} Kernel Optimization},''
  \emph{arXiv preprint arXiv:2511.01884}, 2025. [Online]. Available:
  \url{https://arxiv.org/abs/2511.01884}
\BIBentrySTDinterwordspacing

\bibitem{looprag}
Y.~Zhi, Y.~Cao, J.~Dai, X.~Han, J.~Pu, Q.~Wu, S.~Cheng, and M.~Cai, ``{LOOPRAG:
  Enhancing Loop Transformation Optimization with Retrieval-Augmented Large
  Language Models},'' in \emph{ASPLOS}, 2026.

\bibitem{dist_serve}
Y.~Zhong, S.~Liu, J.~Chen, J.~Hu, Y.~Zhu, X.~Liu, X.~Jin, and H.~Zhang,
  ``{{DistServe}: Disaggregating Prefill and Decoding for Goodput-optimized
  Large Language Model Serving},'' in \emph{OSDI}, 2024.

\bibitem{nanoflow}
K.~Zhu, Y.~Gao, Y.~Zhao, L.~Zhao, G.~Zuo, Y.~Gu, D.~Xie, T.~Tang, Q.~Xu, Z.~Ye,
  K.~Kamahori, C.-Y. Lin, Z.~Wang, S.~Wang, A.~Krishnamurthy, and B.~Kasikci,
  ``{{NanoFlow}: Towards Optimal Large Language Model Serving Throughput},'' in
  \emph{OSDI}, 2025.

\bibitem{megascale_infer}
R.~Zhu, Z.~Jiang, C.~Jin, P.~Wu, C.~A. Stuardo, D.~Wang, X.~Zhang, H.~Zhou,
  H.~Wei, Y.~Cheng, J.~Xiao, X.~Zhang, L.~Liu, H.~Lin, L.-W. Chang, J.~Ye,
  X.~Yu, X.~Liu, X.~Jin, and X.~Liu, ``{MegaScale-Infer: Efficient
  Mixture-of-Experts Model Serving with Disaggregated Expert Parallelism},'' in
  \emph{SIGCOMM}, 2025.

\end{thebibliography}

\end{document}